\documentclass[prl,twocolumn,showpacs,amsfonts,amssymb,amsmath,superscriptaddress]{revtex4-2}

\usepackage{graphicx}
\usepackage{amsmath}
\usepackage{braket}
\usepackage{bm}

	\usepackage{hyperref}		
	\usepackage[figure]{hypcap}		

	\hypersetup{
	   	bookmarks		= false,		
		unicode			= false,	
		pdfborder			= {0 0 0}	
		pdftoolbar			= false,		
		pdfmenubar		= true,		
		pdffitwindow		= false,     	
		pdfstartview		= {FitH},    	
		pdfnewwindow	= true,      	
		colorlinks			= true,		
		linkcolor			= blue,		
		citecolor			= blue,		
		filecolor			= blue,		
		urlcolor			= blue,		
		linktocpage		= true,		
		pdftitle				= {Electronic Raman Scattering in Twistronic Graphene},
		pdfauthor			= {A. Garcia-Ruiz, J.J.P. Thompson, M. Mucha-Kruczynski, V.I. Fal'ko},
		pdfsubject			= {},
		pdfcreator			= {MikTeX},
		pdfproducer		= {Marcin Mucha-Kruczynski},
		pdfkeywords		= {},
}

	\usepackage{xcolor}
	\usepackage[normalem]{ulem}

	\newcommand{\vect}[1]{\boldsymbol{#1}}
	\newcommand{\op}[1]{\hat{{#1}}}
	\newcommand{\Raman}{\omega}
	\newcommand{\X}{\mathrm{X}}
	\newcommand{\Y}{\mathrm{Y}}
	\newcommand{\ttt}{\mathrm{t}}
	\newcommand{\bbb}{\mathrm{b}}		
	\newcommand{\K}{\vect{K}}
	\newcommand{\Kprime}{\vect{K}'}
	\newcommand{\Omegaout}{\tilde{\Omega}}
	\newcommand{\lout}{\tilde{\vect{l}}}
	
	\newcommand{\odots}{\reflectbox{$\ddots$}}

\begin{document}

\title{Electronic Raman Scattering in Twistronic Few-Layer Graphene}
\author{A.~Garc\'{i}a-Ruiz}
\affiliation{Department of Physics, University of Bath, Claverton Down, BA2~7AY, United Kingdom}
\affiliation{National Graphene Institute, University of Manchester, Booth Street East, Manchester, M13 9PL, United Kingdom}
\author{J.~J.~P.~Thompson}
\affiliation{Department of Physics, University of Bath, Claverton Down, BA2~7AY, United Kingdom}
\affiliation{Department of Physics, Chalmers University of Technology, SE-412 96 Gothenburg, Sweden}
\author{M.~Mucha-Kruczy\'{n}ski}
\email{m.mucha-kruczynski@bath.ac.uk}
\affiliation{Department of Physics, University of Bath, Claverton Down, BA2~7AY, United Kingdom}
\affiliation{Centre for Nanoscience and Nanotechnology, University of Bath, Claverton Down, BA2~7AY, United Kingdom}
\author{V.~I.~Fal'ko}
\affiliation{National Graphene Institute, University of Manchester, Booth Street East, Manchester, M13 9PL, United Kingdom}
\affiliation{Department of Physics, University of Manchester, Oxford Road, Manchester, M13 9PL, United Kingdom}

\begin{abstract}

We study electronic contribution to the Raman scattering signals of two-, three- and four-layer graphene with layers at one of the interfaces twisted by a small angle with respect to each other. We find that the Raman spectra of these systems feature two peaks produced by van Hove singularities in moir\'{e} minibands of twistronic graphene, one related to direct hybridization of Dirac states, and the other resulting from band folding caused by moir\'{e} superlattice. The positions of both peaks strongly depend on the twist angle, so that their detection can be used for non-invasive characterization of the twist, even in hBN-encapsulated structures. 

\end{abstract}

\maketitle

Twisted bilayer graphene is a van der Waals heterostructure where the relative twist of constituent atomic planes alters electronic properties of the material \cite{cao_nature_2018_1, cao_nature_2018_2}. A small-angle twist in a bilayer produces a long-period moir\'{e} pattern which generates minibands for electrons with a small moir\'{e} Brillouin zone (mBZ). The minibands and gaps between them strongly depend on twist angle, $\theta$, leading to Mott insulator states and superconductivity for a magic angle, $\theta\approx 1.1^{\circ}$ \cite{cao_nature_2018_1, cao_nature_2018_2, xie_nature_2019, choi_naturephys_2019, codecido_scienceadv_2019, tomarken_prl_2019, jiang_nature_2019, kerelsky_nature_2019, lu_nature_2019, yankowitz_science_2019, stepanov_arxiv_2019}, where the lowest miniband appears to be almost flat. Narrow minibands also appear in $(1+2)$ \cite{chen_arxiv_2020, shi_arxiv_2020} and $(1+1+1)$ \cite{tsai_arxiv_2019} trilayers as well as in $(2+2)$ tetralayers \cite{burg_prl_2019, cao_arxiv_2019, liu_arxiv_2019, rickhaus_arxiv_2020, adak_prb_2020, shen_arxiv_2019, he_arxiv_2020}. Theoretical studies also predict appearance of correlated and topological states in various 'twistronic' graphene stacks \cite{li_arxiv_2019, ma_arxiv_2019, zhang_prb_2019, xu_prl_2018, carr_arxiv_2019, khalaf_prb_2019, mora_prl_2019}, highlighting the need to expand the toolbox of fast and non-invasive methods for measuring twist angles in such structures.  

\begin{figure}[!b]
\includegraphics[width=1\columnwidth]{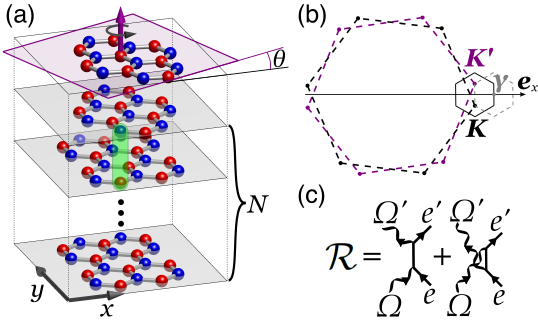}
\caption{ (a) Sketch of $(1+N)$ twistronic graphene. The red and blue balls correspond to the two different sub-lattice sites in each graphene layer and green ellipse represents a dimer bonding leading to direct interlayer coupling. (b) Brillouin zones of the $N$-layer stack (dashed black line) and the top graphene (dashed purple line) with corners $\K$ (bottom layer) and $\Kprime$ (top), respectively, as well as the effective moir\'{e} Brillouin zone, shown both centred around the valleys (solid black line) as the preferred choice in this paper, and with valleys in its corners (dashed grey), centred at $\vect{\gamma}$. (c) Feynman diagrams for the scattering amplitudes $\mathcal{R}$ contributing to the electronic Raman features discussed in this paper. \label{fig:geometry}}
\end{figure}

Over the years, Raman spectroscopy emerged as a powerful tool for the characterization of graphene. Raman scattering with phonons provides information about defects, doping, strain and the number of layers in the film \cite{Ferrari_naturenano_2013}. Twist of graphene layers was found to lead to the resonant enhancement as well as variation of the width and position of various Raman-active phonon modes \cite{ni_prb_2009, havener_nanoletters_2012, carozo_nanoletters_2011, wang_apl_2013, campos-delgado_nanores_2013, he_nanoletters_2012, wu_naturecomms_2014, kim_prl_2012}, however, with a limited accuracy in determining the twist angle. Below, we demonstrate that Raman spectroscopy of the interband electronic excitations (ERS) \cite{kashuba_prb_2009, mucha-kruczynski_prb_2010, riccardi_prl_2016, riccardi_prm_2019, faugeras_prl_2011, garcia-ruiz_prb_2018, kossacki_prb_2011, ponosov_prb_2015, kuhne_prb_2012, faugeras_njp_2012, berciaud_nanoletters_2014, henni_nanoletters_2016, garcia-ruiz_nanoletters_2019} can be used to detect twist-angle-dependent features in the electronic spectrum of twistronic graphene. We study the electronic minibands and electronic contributions to the Raman spectra for few-layer graphene stacks with one of the interfaces between the layers twisted by a small angle, $\theta<2^{\circ}$, Fig.~\ref{fig:geometry}(a). We calculate ERS spectra of twisted bilayers $(1+1)$, trilayers $(1+2)$ and tetralayers [(1+3) and (2+2)] and show that these are formed by transitions from the $n$-th valence to the $n$-th conduction moir\'{e} superlattice (mSL) miniband and feature two spectral peaks. One, at a lower Raman shift, is caused by the resonant hybridization of electronic states of the two few-layer graphene crystals separated by the twisted interface forming the lowest-energy minibands \cite{dossantos_prl_2007, li_naturephys_2010}. Another, higher-energy peak is due to the anti-crossing of bands, backfolded by mSL. Both peaks are related to van Hove singularities in the mSL minibands and involve electronic excitations different from those responsible for the optical absorption \cite{stauber_njp_2013, moon_prb_2013}. We trace the peaks positions as a function of the twist angle and estimate their quantum efficiency, $I\sim 10^{-11}$, to be in the measurable range \cite{kashuba_prb_2009, mucha-kruczynski_prb_2010, riccardi_prl_2016, henni_nanoletters_2016, garcia-ruiz_nanoletters_2019, riccardi_prm_2019, faugeras_prl_2011, kossacki_prb_2011}.

\begin{figure*}[!t]
\includegraphics[width=1\textwidth]{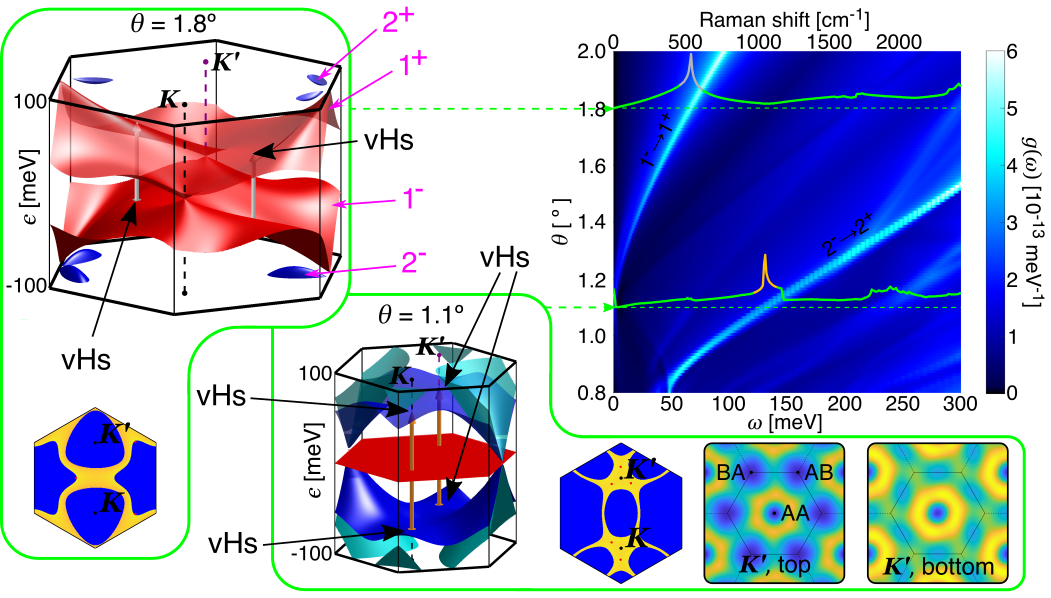}
\caption{Left: Electronic miniband structures of twisted bilayer graphene for $\theta = 1.8^\circ$ and $\theta = 1.1^\circ$ across the corresponding mBZ (solid black hexagons). The dashed lines indicate the positions of the valleys $\K$ (bottom layer) and $\Kprime$ (top) within the mBZ. The pink arrows identify the minibands and the black arrows point to saddle points which give rise to van Hove singularities (vHs). Right: ERS intensity map for $0.8^{\circ}<\theta<2^{\circ}$. The green solid lines show cuts for the angles $\theta = 1.8^\circ$ and $\theta = 1.1^\circ$, indicated with the green dashed lines. The grey and orange peaks mark spectral features corresponding to excitations indicated with vertical arrows of the same colour in the miniband dispersions which involve dispersion saddle points. The hexagonal insets next to band structures show in yellow mBZ regions which contribute to the $1^{-}\rightarrow 1^{+}$ ERS peak for $\theta = 1.8^\circ$ and $2^{-}\rightarrow2^{+}$ peak for $\theta = 1.1^\circ$; red dots indicate positions of dispersion saddle points on the valence side. The additional two insets for $\theta = 1.1^\circ$ depict real space distribution of the saddle points wave function in the top/bottom layers; black solid lines mark boundaries of the moir\'{e} supercell and the letters indicate local stacking.\label{fig:1_1}}
\end{figure*}

To model twistronic graphene, we use a hybrid $k\cdot p$ theory-tight-binding model, where we describe electrons' states in each flake using the $k\cdot p$ expansion around $\pm\K$ and $\pm\Kprime$ Brillouin zone corners of the bottom and anticlockwise rotated (by angle $\theta$) top crystal, respectively [see Fig.~\ref{fig:geometry}(b)] and the interlayer hybridization using tunneling Hamiltonian \cite{dossantos_prl_2007, bistritzer_pnas_2011},
\begin{align}
\op{H} = & \left[\begin{array}{ccc|cccc}
\ddots & \vdots & \vdots & \vdots & \vdots & \vdots & \odots \\
\dots & \op{H}_{+} & \op{T}_{\ttt} & 0 & 0 & 0 & \dots \\
\dots & \op{T}^{\dagger}_{\ttt} & \op{H}_{+} & \op{\mathcal{T}} & 0 & 0 & \dots \\ \hline
\dots & 0 & \op{\mathcal{T}}^{\dagger} & \op{H}_{-} & \op{T}_{\bbb} & 0 & \dots \\
\dots & 0 & 0 & \op{T}^{\dagger}_{\bbb} & \op{H}_{-} & \op{T}_{\bbb'} & \dots \\
\dots & 0 & 0 & 0 & \op{T}^{\dagger}_{\bbb'} & \op{H}_{-} & \dots \\
\odots & \vdots & \vdots & \vdots & \vdots & \vdots & \ddots \\
\end{array}\right], \\
& \op{H}_{\pm} = v\vect{\sigma}_{\xi}\cdot\left(\vect{p}\mp\xi\hbar\tfrac{\theta}{2}K\vect{e}_{y}\right). \nonumber
\end{align}
Here, $\vect{\sigma}_{\xi}=(\xi\sigma_{x},\sigma_{y})$ and $\sigma_{x},\sigma_{y},\sigma_{z}$ are sublattice Pauli matrices for electrons within each monolayer, and the vertical and horizontal lines mark the twisted interface. Momentum $\vect{p}=(p_{x},p_{y})$ is measured from the centre of the valley $\xi\K$ in the bottom flake (with valley index $\xi=\pm 1$), $K=|\vect{K}|$, $\vect{e}_{y}$ is a unit vector along wave vector $y$-axis and $v$ is the monolayer Dirac velocity. To describe $(M\Y+N\X)$ twistronic system, where $M$ and $N$ are the thicknesses of the stack above and below the twisted interface and $\X$ and $\Y$ describe the stacking of the corresponding crystal [$\X=\mathrm{B}$ for Bernal and $\X=\mathrm{R}$ for rhombohedral trilayer, $\X=\mathrm{AB}$ and $\Y=\mathrm{AB}$ or $\Y=\mathrm{BA}$ for $(2+2)$ structures], one needs to truncate $\op{H}$ at $M$ blocks in the top left and $N$ in the bottom right parts. Coupling across the twisted interface is described by \cite{dossantos_prl_2007, bistritzer_pnas_2011}
\begin{align}
&\op{\mathcal{T}}= t \left(
\op{\mathcal{T}}_{0}+
\op{\mathcal{T}}_{1}e^{i\xi\vect{G}_1\cdot\vect{r}}+
\op{\mathcal{T}}_{2}e^{i\xi\vect{G}_2\cdot\vect{r}}\right), \label{eqn:interlayer}\\
&\op{\mathcal{T}}_{n}=
1+\cos\left(\frac{2\pi n}{3}\right)\sigma_x+\xi\sin\left(\frac{2\pi n}{3}\right)\sigma_y, \nonumber
\end{align}
where $\vect{G}_{n}\approx((-1)^{n}\tfrac{1}{2},\tfrac{\sqrt{3}}{2})\sqrt{3}\theta K$ and $t\approx 110$ meV \cite{bistritzer_pnas_2011}. Coupling $\op{\mathcal{T}}$ is responsible for the mSL effects and defines the hexagonal mBZ, chosen as in Fig.~\ref{fig:geometry}(b). Interlayer couplings across the non-twisted interfaces are set using
\begin{align*}
\op{T}(\pm\tfrac{\theta}{2}) & = \begin{bmatrix}
-v_4\op{\pi}^{\dagger}(\pm\frac{\theta}{2}) & v_3\op{\pi}(\pm\frac{\theta}{2}) \\
\gamma_1 & -v_4\op{\pi}^{\dagger}(\pm\frac{\theta}{2})
\end{bmatrix}, \\
\op{\pi}(\pm\tfrac{\theta}{2}) & = \left[\xi p_x+ip_y\mp i\xi\hbar\tfrac{\theta}{2}K\right],
\end{align*}
where $\gamma_{1}=0.39$ eV, $v_{3}\approx 0.11v$, and $v_{4}\approx 0.015v$ \cite{mccann_rpp_2013}. For a $(1+2)$ trilayer, $\op{T}_{\bbb}=\op{T}(-\tfrac{\theta}{2})$. For a $(1+3\mathrm{R})$ stack, $\op{T}_{\bbb}=\op{T}_{\bbb'}=\op{T}(-\tfrac{\theta}{2})$. For a $(1+3\mathrm{B})$ tetralayer, $\op{T}_{\bbb}=\op{T}(-\tfrac{\theta}{2})$ and $\op{T}_{\bbb'}=\op{T}^{\dagger}(-\tfrac{\theta}{2})$. For a $(2\mathrm{AB}+2\mathrm{AB})$ structure, $\op{T}_{\ttt}=\op{T}(\tfrac{\theta}{2})$, $\op{T}_{\bbb}=\op{T}(-\tfrac{\theta}{2})$, while for $(2\mathrm{AB}+2\mathrm{BA})$, $\op{T}_{\bbb}=\op{T}(-\tfrac{\theta}{2})$, $\op{T}^{\dagger}_{\ttt}=\op{T}(\tfrac{\theta}{2})$.

In the electronic Raman scattering, photon of energy $\Omega$ carrying vector potential $\vect{A}=\sqrt{\tfrac{\hbar}{2\epsilon_{0}\Omega}}[\vect{l}e^{i(\vect{q}\cdot\vect{r}-\Omega t)/\hbar}\op{b}_{\vect{q},\vect{l}}+\mathrm{H.c.}]$ ($\op{b}_{\vect{q},\vect{l}}$ annihilates a photon with momentum $\vect{q}$ and polarization $\vect{l}$), arrives at the sample (here, we assume normal incidence of light). This photon scatters to another one with energy $\Omegaout=\Omega-\omega$, momentum $\tilde{\vect{q}}$, and polarization $\lout$, leaving behind an electron-hole pair with energy $\omega$. In contrast to classical plasmas, where the amplitude of such process is controlled by contact interaction, in graphene, the dominating contribution comes from a two-step process, described by the Feynman diagrams shown in Fig.~\ref{fig:geometry}(c). It corresponds to absorption (emission) of a photon with energy $\Omega$ ($\Omegaout$) transferring an electron with momentum $\vect{p}$ from an occupied state in the valence band into a virtual intermediate state (energy is not conserved at this stage), followed by emission (absorption) of a second photon with energy $\Omegaout$ ($\Omega$) \cite{kashuba_prb_2009, mucha-kruczynski_prb_2010}. The amplitude of this process is \cite{garcia-ruiz_nanoletters_2019}
\begin{equation}\label{eqn:amplitude}
\mathcal{R}=i\frac{(e\hbar v)^2}{\epsilon_0\Omega^2}(\vect{l}\times\lout^{*})_z \op{I}_{M+N}\otimes\sigma_z,
\end{equation} 
where $\op{I}_{M+N}$ is a $(M+N)\times(M+N)$ unit matrix. The main contribution to the Raman signal comes from $n^{-}\rightarrow{n^{+}}$ miniband transitions, where $n^{s}$ denotes the $n$-th miniband on the valence ($s=-1$) or conduction ($s=1$) side. The structure of the two-photon coupling vertex in Eq.~\eqref{eqn:amplitude} \cite{kashuba_prb_2009} suggests that, for circularly-polarized photons, both the incoming and outgoing light have the same polarization. In turn, for linearly polarized light, the incoming and outgoing light carries perpendicular polarization (so called linear cross-polarization). To compare, the phonon $G$-line is observed in both parallel and perpendicular linear polarization geometries whereas for circular polarization the incoming and outgoing photons have opposite polarization \cite{basko_prb_2008, faugeras_prl_2011, kossacki_prb_2011, ponosov_prb_2015, kuhne_prb_2012, faugeras_njp_2012}. The overall lineshape $g(\Raman)$ of inelastic photon scattering with Raman shift, $\Raman$, is given by,
\begin{align}\label{eqn:ers}
g(\Raman) & =\frac{1}{c}\int\frac{\mathrm{d}\tilde{\vect{q}}\,w(\Raman)}{(2\pi\hbar)^3}\delta(\Omegaout-c|\tilde{\vect{q}}|)\approx\frac{\Omega^{2}}{(2\pi\hbar)^{3} c^4}w(\Raman),\nonumber\\
w(\Raman) & =\frac{2}{\pi\hbar^{3}}\sum_{n^{s},m^{s'}}\int\mathrm{d}\vect{p}\left|\Braket{\vect{p},m^{s'}|\mathcal{R}|\vect{p},n^{s}}\right|^2 \\
&  \times f_{\vect{p},n^{s}}(1-f_{\vect{p},m^{s'}})\delta(\epsilon_{\vect{p},m^{s'}}-\epsilon_{\vect{p},n^{s}}-\Raman),\nonumber
\end{align}
where $f_{\vect{p},n^{s}}$ is the occupation factor of a state $\ket{\vect{p},n^{s}}$ with momentum $\vect{p}$ and energy $\epsilon_{\vect{p},n^{s}}$ in the band $n^{s}$. 

\begin{figure}[!t]
\includegraphics[width=1\columnwidth]{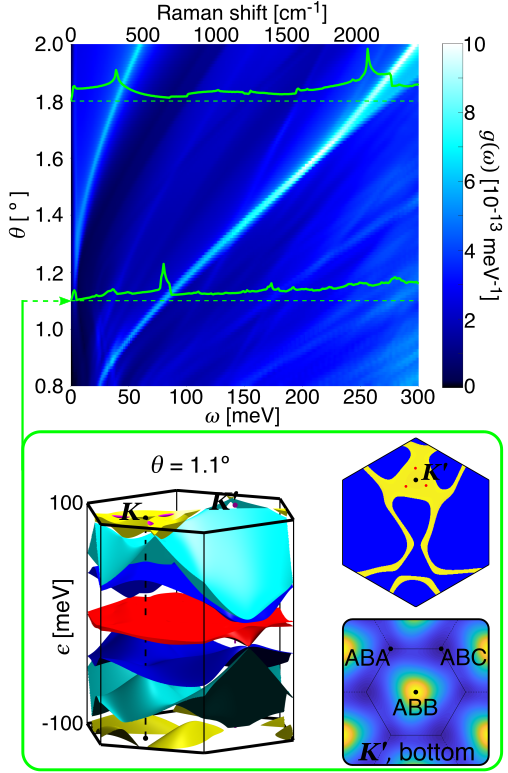}
\caption{Top: ERS intensity map for (1+2) graphene. The green dashed lines indicate twist angles $\theta=1.8^{\circ}$ and $\theta=1.1^{\circ}$ and the green solid lines show the ERS spectra for these angles. Bottom: Miniband structure for the $(1+2)$ stack for $\theta=1.1^{\circ}$. The top inset shows in yellow the mBZ regions which contribute to the $2^{-}\rightarrow 2^{+}$ ERS peak; red dots indicate positions of dispersion saddle points on the valence side. The saddle points wave function is predominantly located in the bottom layer and its real space distribution is presented in the bottom inset; black solid lines mark boundaries of the moir\'{e} supercell and the letters indicate local stacking (from the bottom layer to the top). \label{fig:1_2}}
\end{figure}

\begin{figure*}[!t]
\includegraphics[width=1\textwidth]{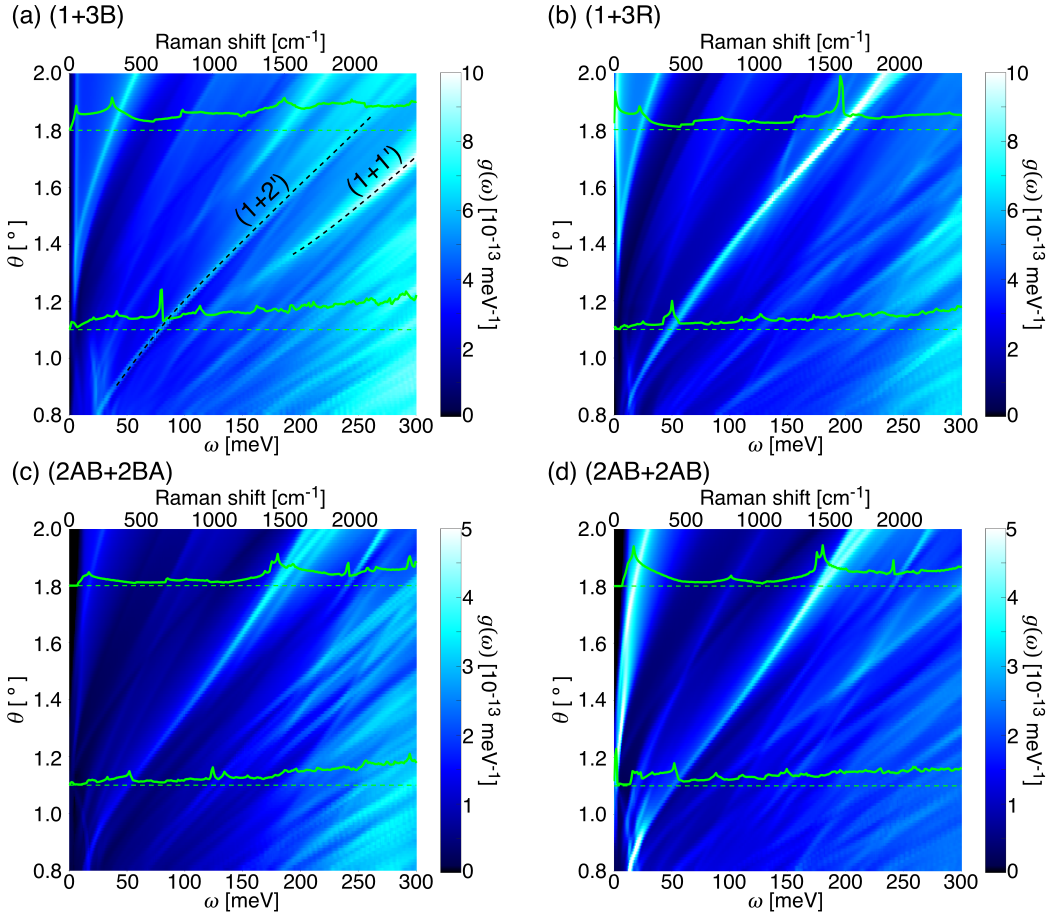}
\caption{Comparison of ERS intensity maps for (a) $(1+3\mathrm{B})$, (b) $(1+3\mathrm{R})$, (c) $(2\mathrm{AB}+2\mathrm{BA})$, and (d) $(2\mathrm{AB}+2\mathrm{AB})$ tetralayers. Green solid lines show the Raman spectra for twist angles $\theta=1.8^{\circ}$ and $\theta=1.1^{\circ}$. \label{fig:1_3}}
\end{figure*}

In Fig. ~\ref{fig:1_1} we present the ERS intensity map for a $(1+1)$ twisted bilayer graphene with $0.8^{\circ}<\theta<2^{\circ}$, neglecting its atomic reconstruction \cite{yoo_naturemat_2019}. Two bright features stand out in this plot, corresponding to transitions between the minibands $1^{-}\rightarrow 1^{+}$ and $2^{-}\rightarrow 2^{+}$. For two selected cuts at  $\theta=1.8^{\circ}$ and $\theta=1.1^{\circ}$, the peaks come from transitions indicated by arrows in the corresponding minibands plots. The first peak, $1^{-}\rightarrow{1^{+}}$, is due to transitions from (or to) flat regions of electronic dispersion resulting from hybridization of the Dirac cones of the two layers (the Dirac points can be identified in the dispersion for $\theta=1.8^{\circ}$ as touching points of the minibands shown in red), which are responsible for a van Hove singularity (vHs) in the density of states \cite{dossantos_prl_2007, li_naturephys_2010} for larger twist angles. Note that due to the chiral nature of graphene electrons, the vHs is not positioned on the line connecting the cones but is shifted off it in the opposite directions in the conduction and valence bands (as indicated by black arrows in Fig.~\ref{fig:1_1} for $\theta=1.8^{\circ}$). The inset below the $\theta=1.8^{\circ}$ miniband structure highlights in yellow parts of the mBZ that contribute to the $1^{-}\rightarrow{1^{+}}$ peak. As the twist angle is decreased, this ERS peak moves to lower energies. The second peak, $2^{-}\rightarrow{2^{+}}$, is due to the transitions between the flat regions of the second valence and conduction minibands, indicated by orange arrows in the $\theta=1.1^{\circ}$ miniband plot. Its intensity comes from the mBZ section painted in yellow in the left-most inset below the ERS map. The other two insets show the real-space distribution of the saddle point states across the moir\'{e} supercell in the top and bottom monolayers. 

In Fig.~\ref{fig:1_2}, we show the ERS intensity map for $(1+2)$ twistronic graphene and an exemplary miniband structure for $\theta=1.1^{\circ}$. Similarly to $(1+1)$ graphene, the dominant contributions come from the $1^{-}\rightarrow 1^{+}$ and $2^{-}\rightarrow 2^{+}$ electronic transitions. The two peaks also have the same origins: the first one is due to direct hybridization of the monolayer and bilayer states while the second is formed by states backfolded by moir\'{e} superlattice. However, for a given twist angle, the peaks appear at lower Raman shifts than in the $(1+1)$ structure. This is because the unperturbed dispersion in bilayer is flatter than in a monolayer -- as a consequence, the anti-crossings vHs form at lower energies than in a $(1+1)$ stack. 

In Fig.~\ref{fig:1_3}(a) and (b), we plot ERS maps of two monolayer-on-trilayer structures, $(1+3\mathrm{B})$ and $(1+3\mathrm{R})$. Bernal-stacked trilayer graphene hosts both a bilayer- and monolayer-like low-energy bands \cite{koshino_prb_2010}. Both of these hybridize with the states of the top monolayer to form the first miniband and contribute to the $1^{-}\rightarrow 1^{+}$ ERS peak. The next peak, marked as $(1+2')$ in Fig.~\ref{fig:1_3}(a), is due to the anti-crossing of backfolded bilayer-like and top monolayer bands. Another peak, $(1+1')$, is due to an anti-crossing of the backfolded top monolayer band and monolayer-like band of the trilayer. In contrast, rhombohedral trilayers only host one low-energy band, with a flat dispersion in the vicinity of the valley centre \cite{koshino_prb_2009}. This low-energy band is localised on the top and bottom surfaces of the crystal (representing surface states generic for rhombohedral graphitic films \cite{guinea_prb_2006, slizovskiy_communphys_2019, henck_prb_2018}) and is significantly affected by mSL, leading to a pair of clear spectral features, Fig.~\ref{fig:1_3}(b), as in $(1+1)$ and $(1+2)$ twistronic graphenes (in the Supplemental Material, we describe a simple model that can be used to determine approximate positions of these ERS peaks \cite{SM}).

In Fig.~\ref{fig:1_3}(c) and (d), we present ERS spectra for two structurally inequivalent double-bilayer stacks, $(2\mathrm{AB}+2\mathrm{BA})$ and $(2\mathrm{AB}+2\mathrm{AB})$. The details of their lattice structure and exemplary miniband dispersions are shown in the Supplemental Material \cite{SM}. The miniband spectra of AB/AB and AB/BA tetralayers are nearly identical, leading to the $1^{-}\rightarrow1^{+}$ and $2^{-}\rightarrow2^{+}$ ERS peaks at the same energies. However, due to  the difference in the wave functions, the intensity of the $1^{-}\rightarrow1^{+}$ is much higher for $(2\mathrm{AB}+2\mathrm{AB})$ than for $(2\mathrm{AB}+2\mathrm{BA})$ tetralayers (see Supplemental Material \cite{SM}).

Overall, electronic Raman scattering spectra of twistronic graphene contain characteristic features related to the van Hove singularities of moir\'{e} superlattice minibands, which depend on the twist angle between the layers. Currently, accurate determination of the twist angle, especially in the small-angle regime, requires time-consuming microscopic investigations of the moir\'{e} pattern, or magneto-transport measurements at cryogenic temperatures. We suggest that, based on our results, calibration of the positions of the Raman features in the structures with known $\theta$ would allow to identify the orientations of the component crystals in other samples, enabling a non-invasive method for measuring the twist angle, even in structures encapsulated with other materials, where the graphene/graphene moir\'{e} pattern is not directly accessible for tunnelling spectroscopy studies \cite{cao_nature_2018_1, cao_nature_2018_2, codecido_scienceadv_2019, tomarken_prl_2019, lu_nature_2019, yankowitz_science_2019, stepanov_arxiv_2019, chen_arxiv_2020, shi_arxiv_2020, tsai_arxiv_2019, burg_prl_2019, cao_arxiv_2019, liu_arxiv_2019, rickhaus_arxiv_2020, adak_prb_2020, shen_arxiv_2019, he_arxiv_2020}.

\section{\label{sec:acknowledgements}Acknowledgements}

This work has been supported by the UK Engineering and Physical Sciences Research Council (EPSRC) through the Centre for Doctoral Training in Condensed Matter Physics (CDT-CMP), Grant No.~EP/L015544/1, EPSRC Grants EP/M507982/1, EP/S019367/1, EP/P026850/1 and EP/N010345/1, the European Graphene Flagship Core3 project (EC Grant Agreement No. 881603), European Research Council Synergy Grant Hetero2D and Lloyd's Register Foundation Nanotechnology Programme. M.~M.-K.~also acknowledges funding through the University of Bath International Research Funding Scheme.

\end{document}


\beginsupplement
	
\title{Supplemental Material: \\ Electronic Raman Scattering in Twistronic Few-Layer Graphene}

\author{A.~Garc\'{i}a-Ruiz}
\affiliation{Department of Physics, University of Bath, Claverton Down, BA2~7AY, United Kingdom}
\affiliation{National Graphene Institute, University of Manchester, Booth Street East, Manchester, M13 9PL, United Kingdom}
\author{J.~J.~P.~Thompson}
\affiliation{Department of Physics, University of Bath, Claverton Down, BA2~7AY, United Kingdom}
\affiliation{Department of Physics, Chalmers University of Technology, SE-412 96 Gothenburg, Sweden}
\author{M.~Mucha-Kruczy\'{n}ski}
\email{m.mucha-kruczynski@bath.ac.uk}
\affiliation{Department of Physics, University of Bath, Claverton Down, BA2~7AY, United Kingdom}
\affiliation{Centre for Nanoscience and Nanotechnology, University of Bath, Claverton Down, BA2~7AY, United Kingdom}
\author{V.~I.~Fal'ko}
\affiliation{National Graphene Institute, University of Manchester, Booth Street East, Manchester, M13 9PL, United Kingdom}
\affiliation{Department of Physics, University of Manchester, Oxford Road, Manchester, M13 9PL, United Kingdom}

\maketitle

\tableofcontents

\newpage

\section{Electronic minibands of four-layer twistronic graphene}

\begin{figure}[b]
\includegraphics[width=1.0\columnwidth]{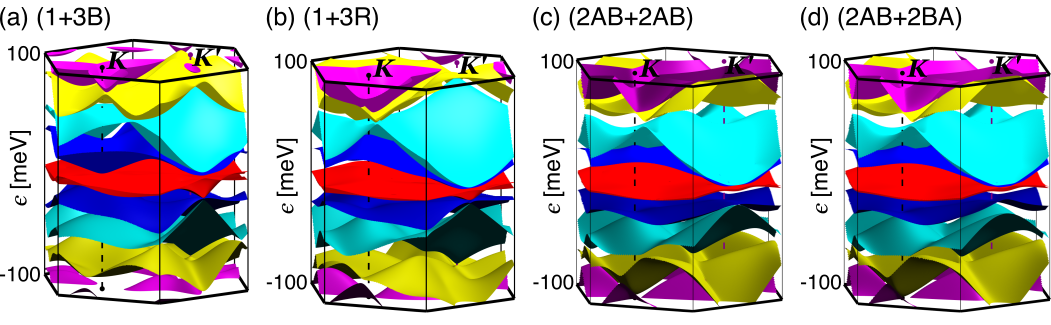}
\caption{Miniband structures of four-layer twistronic graphene for $\theta=1.1^{\circ}$: (a) (1+3B), (b) (1+3R), (c) (2AB+2AB) and (d) (2AB+2BA). The dashed black and purple lines indicate the positions of the valleys $\K$ (bottom layer) and $\Kprime$ (top layer) within the mBZ. \label{fig:minibands}}
\end{figure}

In Fig.~\ref{fig:minibands}, we compare the miniband structures for all tetralayer twistronic graphenes with a single twisted interface and $\theta=1.1^{\circ}$. Notice the flatness of the first miniband for the (1+3R) and (2+2) structures, leading to the $1^{-}\rightarrow 1^{+}$ ERS peak appearing at the lowest energies amongst all of the investigated structures, as seen in Fig.~4 of the main text. While in the (1+3R) structure this is related to the presence and dispersion of the surface state, in double bilayer stacks the flatness is generated by coupling of two quasi-quadratic dispersions, yielding minigaps and van Hove singularities at lower energies than in the presence of linear dispersion contributed in $(1+N)$ structures by the top monolayer. Importantly, as we discuss below, the low-energy electronic band structure of the (2AB+2AB) and (2AB+2BA) stacks is essentially the same.

\begin{figure}[t]
\includegraphics[width=1.0\columnwidth]{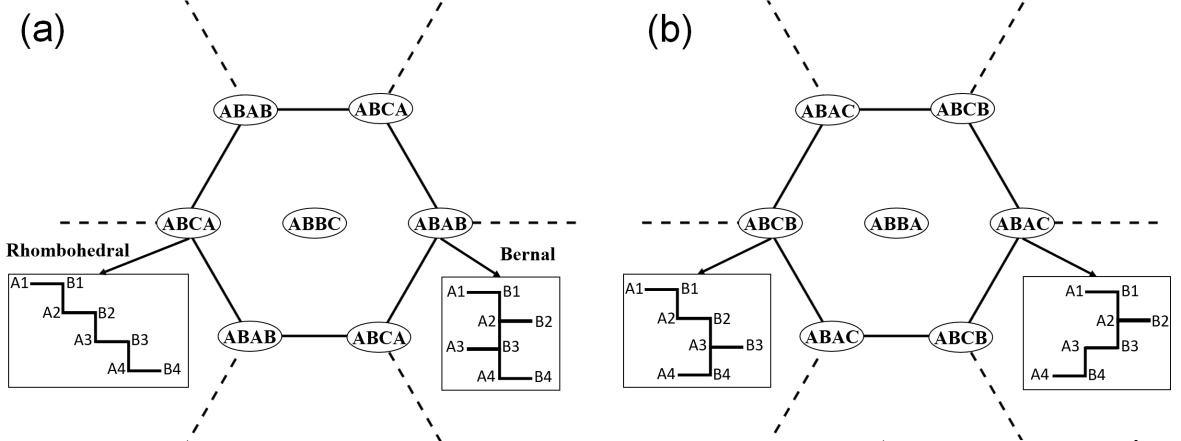}
\caption{Comparison of local stackings within the moir\'{e} unit cell (solid black hexagon) for the (a) (2AB+2AB) and (b) (2AB+2BA) (right) structures. Insets show schematic side-view representation of local stackings at positions indicated by the arrows.\label{fig:doubleblg}}
\end{figure}

For graphene stacks, twist induces periodic variation of the offset of the lattices of the adjacent layers (stackings). At the twisted interface, a shift of one layer with respect to the other by $\delta$ leads to the shift of the offset pattern (superlattice) by $\delta/\theta$ (for example, see in \cite{cosma_faraday_2014}), which is equivalent to a change of co-ordinates for the whole system. As a result, for twistronic structures an additional shear displacement (at the twisted interface) of one crystal with respect to the other would not change moir\'{e} superlattice minibands. Equivalently, stacking at the twisted interface prior to twisting and position of the axis of rotation do not influence the moir\'{e} pattern: for example, both ideal ABAB and ABAC tetralayers lead to the (1+3B) structure upon introducing a twist between the last two layers; however, (1+3B) and (1+3R) are inequivalent because of the different stackings in the trilayer crystal. In the case of (2+2) structures, for $\theta=0$ two different inequivalent stackings built out of two Bernal bilayers are possible. The first one, ABAB, is such that a perfect Bernal stacking is maintained across the flake. The second is obtained from the first by a $180^{\circ}$-degree rotation of the top bilayer about an axis passing through one of the sites at which carbon atoms of the four layers form a vertical chain. Such a rotation leads to the ABBA structure which, on top of the inversion symmetry present in ABAB, also possesses reflection symmetry. As shown in Fig.~\ref{fig:doubleblg}, the two structures remain inequivalent in the presence of a twist between the bilayer blocks as the local stackings present within the moir\'{e} unit cell are different in both cases.

The $180^{\circ}$-degree rotation of the top bilayer in the AB/BA structure as compared to AB/AB is reflected in the corresponding Hamiltonians of the twisted structures,
\begin{align*}
\op{H}_{2\mathrm{AB}+2\mathrm{AB}}= &
\begin{bmatrix}
\op{H}_{+} & \op{T}(\tfrac{\theta}{2}) & 0 & 0 \\
\op{T}^{\dagger}(\tfrac{\theta}{2}) & \op{H}_{+} & \op{\mathcal{T}} & 0 \\
0 & \op{\mathcal{T}}^{\dagger} & \op{H}_{-} & \op{T}(-\tfrac{\theta}{2}) \\
0 & 0 & \op{T}^{\dagger}(-\tfrac{\theta}{2}) & \op{H}_{-}
\end{bmatrix}, \\
\op{H}_{2\mathrm{AB}+2\mathrm{BA}}= &
\begin{bmatrix}
\op{H}_{+} & \op{T}^{\dagger}(\tfrac{\theta}{2}) & 0 & 0 \\
\op{T}(\tfrac{\theta}{2}) & \op{H}_{+} & \op{\mathcal{T}} & 0 \\
0 & \op{\mathcal{T}}^{\dagger} & \op{H}_{-} & \op{T}(-\tfrac{\theta}{2}) \\
0 & 0 & \op{T}^{\dagger}(-\tfrac{\theta}{2}) & \op{H}_{-}
\end{bmatrix}.
\end{align*}
In spite of these structural differences, band structures of the two twisted double bilayers are almost identical. This similarity can be explained by working with effective low-energy Hamiltonians for Bernal bilayers \cite{mccann_prl_2006},
\begin{align}
\op{\tilde{H}}_{2}(\pm\tfrac{\theta}{2})=
\begin{bmatrix}
0 & \frac{v^2}{\gamma_{1}}\left(\op{\pi}^{2}(\pm\tfrac{\theta}{2})\right)^{\dagger} \\ \frac{v^2}{\gamma_{1}}\op{\pi}^{2}(\pm\tfrac{\theta}{2}) & 0
\end{bmatrix}.
\end{align}
The bottom bilayer is arranged the same way for both configuration, but the for the top bilayer, the dimer sites are in different positions. Accordingly, the effective Hamiltonian in each case is
\begin{align*}
\op{\tilde{H}}_{2\mathrm{AB}+2\mathrm{AB}} & =
\begin{bmatrix}
0 & \frac{v^2}{\gamma_{1}}\left(\op{\pi}^{2}(\tfrac{\theta}{2})\right)^{\dagger} & 0 & 0 \\
\frac{v^2}{\gamma_{1}}\op{\pi}^{2}(\tfrac{\theta}{2}) & 0 & \op{\mathcal{T}}_{\mathrm{BA}}(\theta) & 0 \\
0 & \op{\mathcal{T}}_{\mathrm{BA}}^{\dagger}(\theta) & 0 & \frac{v^2}{\gamma_{1}}\left(\op{\pi}^{2}(-\tfrac{\theta}{2})\right)^{\dagger} \\
0 & 0 & \frac{v^2}{\gamma_{1}}\op{\pi}^{2}(-\tfrac{\theta}{2}) & 0
\end{bmatrix}, \\
\op{\tilde{H}}_{2\mathrm{AB}+2\mathrm{BA}} & =
\begin{bmatrix}
0 & \frac{v^2}{\gamma_{1}}\left(\op{\pi}^{2}(\tfrac{\theta}{2})\right)^{\dagger} & 0 & 0 \\
\frac{v^2}{\gamma_{1}}\op{\pi}^{2}(\tfrac{\theta}{2}) & 0 & \op{\mathcal{T}}_{\mathrm{AA}}(\theta) & 0 \\
0 & \op{\mathcal{T}}_{\mathrm{AA}}^{\dagger}(\theta) & 0 & \frac{v^2}{\gamma_{1}}\left(\op{\pi}^{2}(-\tfrac{\theta}{2})\right)^{\dagger} \\
0 & 0 & \frac{v^2}{\gamma_{1}}\op{\pi}^{2}(-\tfrac{\theta}{2}) & 0
\end{bmatrix},
\end{align*}
with the interlayer coupling at the twisted interface coupling only one of the atomic sites in each of the bilayers,
\begin{align*}
\op{\mathcal{T}}_{BA} & = t \left( 1 + e^{i\frac{2\pi}{3}}e^{i\xi\vect{G}_1\cdot\vect{r}} + e^{-i\frac{2\pi}{3}}e^{i\xi\vect{G}_2\cdot\boldsymbol{r}} \right), \\
\op{\mathcal{T}}_{AA} & = t \left( 1 + e^{-i\frac{2\pi}{3}}e^{i\xi\vect{G}_1\cdot\vect{r}} + e^{+i\frac{2\pi}{3}}e^{i\xi\vect{G}_2\cdot\boldsymbol{r}} \right).
\end{align*}
Note that the effective Hamiltonians above are equivalent, as after a translation $\vect{r}\to\vect{r}+(\vect{a}_1-\vect{a}_2)/3$, with $\vect{a}_1$ and $\vect{a}_2$ the unit vectors of the moir\'{e} superlattice, $\op{\mathcal{T}}_{BA}$ transforms into $\op{\mathcal{T}}_{AA}$. This, in turn, might suggest that the ERS spectra of the two structures are very similar. However, while the positions of the ERS peaks are prescribed by the shape of miniband dispersions which are very similar, this alone does not guarantee the same Raman intensities for both structures. In particular, notice that in Fig.~4 of the main text, the intensity of the $1^{-}\rightarrow 1^{+}$ ERS peak is much higher for the AB/AB structure than for AB/BA. This is related to the rotation of the top bilayer crystal by $180^{\circ}$ in the latter stack which affects the Raman scattering amplitude, 
\begin{align}
\mathcal{R}^{\mathrm{AB/AB}}\propto
\begin{bmatrix}
\sigma_z & 0 \\ 0 & \sigma_z
\end{bmatrix},\,\,\,\,
\mathcal{R}^{\mathrm{AB/BA}}\propto
\begin{bmatrix}
-\sigma_z & 0 \\ 0 & \sigma_z
\end{bmatrix}.
\end{align}

\section{Approximate positions of the ERS peaks}

We estimate the positions of the $1^{-}\rightarrow 1^{+}$ and $2^{-}\rightarrow 2^{+}$ ERS peaks in twistronic graphene by using a simple model to approximate the positions of the corresponding van Hove singularities in the miniband dispersions. We consider crossings of the unperturbed dispersions which generate anticrossings and open minigaps in the electronic spectrum and assume that the vHs are generated by electronic states immediately below the anticrossing. For the first ERS, peak, $1^{-}\rightarrow 1^{+}$, we consider direct hybridization of states in between the valleys $\K$ and $\Kprime$ separated by $\deltaK\approx\theta K$. For a structure $(M+N)$, where $M$ and $N$ correspond to the thicknesses of the two graphene crystals meeting at a twisted interface, these states cross at the energy $\epsilon_{M+N}^{(1)}$ and are coupled to each other with strength $t^{(1)}_{M+N}$. For the second peak, $2^{-}\rightarrow 2^{+}$, we consider the lowest crossing of states backfolded into the moir\'{e} Brillouin zone which is equivalent to studying crossings of dispersions shifted from the valley centres $\K$ and $\Kprime$ by one of the shortest moir\'{e} reciprocal vectors. Such valley centre replicas are separated by the distance $2\deltaK\approx 2\theta K$, with the bands crossing at the energy $\epsilon_{M+N}^{(2)}$ and coupled with strength $t^{(2)}_{M+N}$. 

Within this simple model, perturbation theory yields a local Hamiltonian for two interacting bands,
\begin{align}
\op{H}^{(n)}_{M+N} = \begin{bmatrix} \epsilon_{M+N}^{(n)} & t^{(n)}_{M+N} \\ (t^{(n)}_{M+N})^{*} & \epsilon_{M+N}^{(n)} \end{bmatrix}.
\end{align}
Introduction of the coupling leads to repulsion of the states which shift to energies $\epsilon_{M+N}^{(n)}+|t^{(n)}_{M+N}|$ and $\epsilon_{M+N}^{(n)}-|t^{(n)}_{M+N}|$. We equate the latter with the positions of the vHs in the dispersion and assume electron-hole symmetry of the miniband structure (which holds for $v_{4}=0$) so that the position of the ERS peak corresponding to the $n^{-}\rightarrow n^{+}$, $n=1,2$, transition is
\begin{align}
\epsilon^{n^{-}\rightarrow n^{+}}_{M+N}=2\left[\epsilon_{M+N}^{(n)} - \lvert t^{(n)}_{M+N}\rvert\right].
\end{align}

Below, we discuss the energies $\epsilon_{M+N}^{(n)}$ and $t^{(n)}_{M+N}$ for all of the structures studied in the main text.
\begin{enumerate}[i.]
\item (1+1)

Crossing of two identical linear dispersions with velocities $v$, Dirac points at the same energy and separated by wave vector $\vect{q}$ occurs at the energy $\tfrac{1}{2}\hbar vq$. At the point of degeneracy, graphene eigenstates are coupled by a matrix $t\mathcal{T}_{1}=t(I_{2}+\sigma_{x})$. Taking into account the chiral form of the eigenstates \cite{castro_neto_rmp_2009}, we obtain
\begin{align*}
\epsilon_{1+1}^{(1)}&=\tfrac{1}{2}\hbar v\theta K, & t^{(1)}_{1+1} & = \frac{1}{2}t(1+2ie^{i\frac{\theta}{2}}-e^{-i\theta}) \approx it , \\
\epsilon_{1+1}^{(2)}&=\hbar v\theta K, & t^{(2)}_{1+1}& = t^{(1)}_{1+1} \approx it.
\end{align*}
   
\item (1+2)

To determine the energy position of a crossing between dispersions of a monolayer and bilayer, we use the following form of bilayer graphene low-energy dispersion \cite{mccann_prl_2006},
\begin{align*}
\epsilon = \frac{1}{2}\left[ \sqrt{\gamma_{1}^{2}+4(\hbar vp)^{2}} - \gamma_{1} \right],
\end{align*}
where wave vector distance $p$ is measured from the centre of the valley and we negelcted the trigonal warping caused by $v_{3}$. The relevant crossings with the linear dispersion of monolayer occur at
\begin{align*}
\epsilon_{1+2}^{(1)} = \frac{(\hbar v\theta K)^{2}}{\gamma_{1}+2\hbar v\theta K}, \,\,\,\, \epsilon_{1+2}^{(2)} = \frac{4(\hbar v\theta K)^{2}}{\gamma_{1}+4\hbar v\theta K}.
\end{align*}
In turn, the coupling between the energy eigenstates is
\begin{align*}
t^{(1)}_{1+2}& = t^{(2)}_{1+2} \approx -\frac{1}{2}t(1+i).
\end{align*}

\item (2+2)

For two identical bilayer dispersions, the relevant crossing points are given by
\begin{align*}
\epsilon_{2+2}^{(1)} &= \frac{1}{2}\left(\sqrt{\gamma_{1}^{2}+(\hbar v\theta K)^2} - \gamma_{1}\right),\\
\epsilon_{2+2}^{(2)} &= \frac{1}{2}\left(\sqrt{\gamma_{1}^{2}+4(\hbar v\theta K)^2} - \gamma_{1}\right),
\end{align*}
where we used the expression for bilayer graphene dispersion from above. The direct coupling due to moir\'{e} is $t^{(1)}_{2+2} = t^{(2)}_{2+2} \approx -\tfrac{1}{2}t$.

\item (1+3B)

In the absence of couplings other than in-plane hoppings and direct interlayer coupling $\gamma_{1}$, electronic dispersion of Bernal trilayer can be decomposed into a monolayer-like linear band with the same dispersion, $\epsilon=\hbar vp$, and a bilayer-like band with dispersion modified by an effective interlayer-like coupling $\sqrt{2}\gamma_{1}$ \cite{koshino_prb_2009_ABA}. The bilayer-like bands crosses the linear band of twisted monolayer at lower energies than the monolayer-like band and formulae used for the (1+2) structure, with $\gamma_{1}\to\sqrt{2}\gamma_{1}$ and $t^{(1)}_{1+3B} = t^{(2)}_{1+3B} \approx -\frac{1}{2\sqrt{2}}t(1+i)$, can be applied here to obtain the positions of $1^{-}\rightarrow 1^{+}$ and $2^{-}\rightarrow 2^{+}$ ERS peaks. However, note that the simple model used here does not capture the contribution of the monolayer-like states to the position of the $1^{-}\rightarrow 1^{+}$ peak. For small momenta, these states are close in energy to the bilayer-like bands and equally strongly coupled to the linear band of twisted monolayer, hybridizing with the latter and influencing the position of the $1^{-}\rightarrow 1^{+}$ peak. Finally, the monolayer-like band generates a crossing with the linear band of twisted monolayer, to which it is coupled with the strength $\tfrac{i}{\sqrt{2}}t$. Interaction of these two states leads to an ERS peak marked as (1+1') in Fig.~4(a). 

\item (1+3R)

While an effective model can be produced to describe the dispersion of the surface state \cite{koshino_prb_2009_ABC}, its leading term, $\propto\tfrac{v^{3}}{\gamma_{1}^{2}}p^{3}$, does not lead to a simple analytic solution for its crossing with the linear dispersion of the twisted monolayer. Because of this, we look for the crossing point numerically, using the full Hamiltonian for rhombohedral trilayer, albeit with $v_{4}=0$ to preserve the electron-hole symmetry of the band structure. The couplings between the monolayer and trilayer states are the same as used in the case of the (1+2) and (1+3B) structures, $|t^{(1)}_{2+2}| = |t^{(2)}_{2+2}| \approx |t/\sqrt{2}|$.

\end{enumerate} 

\begin{figure}[t]
\includegraphics[width=0.5\columnwidth]{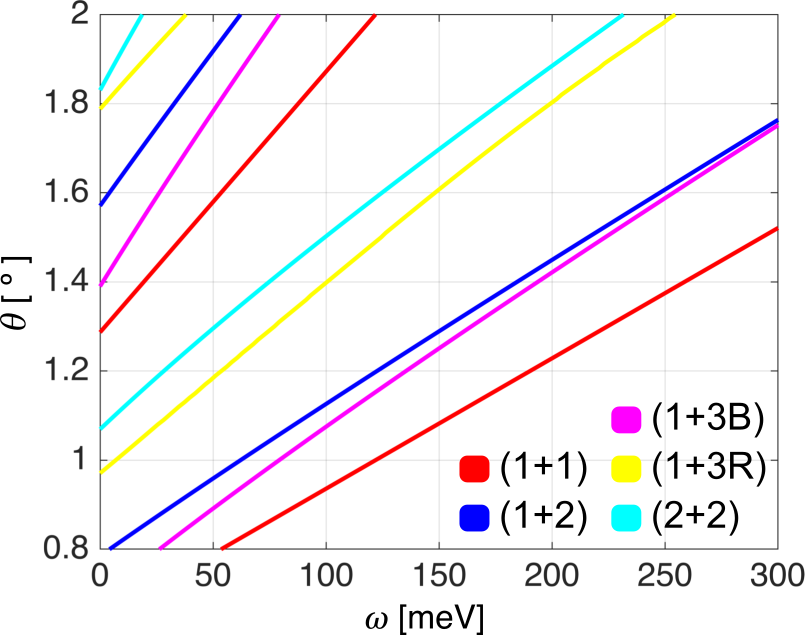}
\caption{Approximate positions of the ERS peaks in twistronic graphene as a function of the twist angle $\theta$. Transitions for the (1+1), (1+2), (1+3B), (1+3R) and (2+2) are shown in red, blue, magenta, yellow and cyan, respectively. \label{fig:approxERS}}
\end{figure}

The results of our approximate model for the positions of the ERS transitions for all the structures discussed in the main text are shown in Fig.~\ref{fig:approxERS}. Two lines are shown for each twistronic stack, with the one at low energies corresponding to the $1^{-}\rightarrow 1^{+}$ transition and the other to the $2^{-}\rightarrow 2^{+}$ transition. Note that the Hamiltonian $\op{H}^{(1)}_{M+N}$ is unable to capture flattening of the first minband in the vicinity of the magic angle which requires taking into account more states coupled by the moir\'{e} supperlattice \cite{bistritzer_pnas_2011}. This leads to positions of the first ERS peak going to zero at angles above the magic angle for the corresponding structure, below which the approximation cannot be used. However, the large-angle limit of the $1^{-}\rightarrow 1^{+}$ peak position as well as position of the $2^{-}\rightarrow 2^{+}$ peak as a function of $\theta$ are well captured by this model.